\documentclass[11pt,twoside]{article}
\usepackage{asp2014}

\aspSuppressVolSlug
\resetcounters

\begin{document}

\title{SPEX: High-Resolution Spectral Modeling \\ and Fitting for X-ray Astronomy}

\author{Jelle~de~Plaa,$^1$ Jelle~S.~Kaastra,$^{1,2}$ Liyi Gu,$^{3}$ Junjie Mao,$^{4}$ and Ton Raassen$^{1}$}
\affil{$^1$SRON Netherlands Institute for Space Research, Utrecht, The Netherlands; \email{j.de.plaa@sron.nl}}
\affil{$^2$Leiden University, Leiden, The Netherlands}
\affil{$^3$RIKEN High Energy Astrophysics Laboratory, Wako, Saitama, Japan }
\affil{$^4$University of Strathclyde, Glasgow, UK}

\paperauthor{Jelle de Plaa}{j.de.plaa@sron.nl}{0000-0002-2697-7106}{SRON}{}{Utrecht}{}{3584CA}{The Netherlands}
\paperauthor{Jelle S. Kaastra}{j.s.kaastra@sron.nl}{0000-0001-5540-2822}{SRON}{}{Utrecht}{}{3584CA}{The Netherlands}
\paperauthor{Liyi Gu}{l.gu@sron.nl}{0000-0001-9911-7038}{RIKEN}{High Energy Astrophysics Laboratory}{Wako}{Saitama}{351-0198}{Japan}
\paperauthor{Junjie Mao}{junjie.mao@strath.ac.uk}{0000-0001-7557-9713}{University of Strathclyde}{Department of Physics}{Glasgow}{}{G4 0NG}{UK}
\paperauthor{Ton Raassen}{a.j.j.raassen@sron.nl}{}{SRON}{}{Utrecht}{}{3584CA}{The Netherlands}




\begin{abstract}
We present the SPEX software package for modeling and fitting X-ray
spectra. Our group has developed spectral models, atomic data and 
code for X-ray applications since the 1970's. Since the 1990's these
are further developed in the public SPEX package. In the last decades,
X-ray spectroscopy has been revolutionized by the high-resolution 
grating spectrometers aboard XMM-Newton and Chandra. Currently, new
high-resolution detectors aboard the Hitomi mission, and future missions,
like XRISM and Athena, will provide another major step forward in
spectral resolution. This poses challenges for us to increase the atomic
database substantially, while keeping model calculation times
short. In this paper, we summarize our efforts to prepare the SPEX 
package for the next generation of X-ray observatories.
\end{abstract}

\section{Introduction}

Plasmas, gases and dust in the Universe can show distinct spectral features 
in the X-ray band. With the advance of space-based X-ray observatories that 
started in the 1970's, there has been an increasing demand for spectral models 
that can be used to fit the observed spectra. What started with a code to
model the solar X-ray spectrum \citep{mewe1972} grew over a few decades 
into a multi-purpose spectral fitting code called SPEX\footnote{\url{https://www.sron.nl/astrophysics-spex}} 
\citep{kaastra1996}.
The package contains models for, for example, optically thin plasmas in collisional
ionization equilibrium, plasmas in photo-ionization equilibrium, and charge-exchange 
emission. These models are all using the same atomic database, which is 
built-up in parallel with the software package.

SPEX has a similar purpose as the other well known X-ray spectral fitting package 
called Xspec \citep{arnaud1996}. The main difference is that Xspec acts as a spectral
fitting platform with community provided models, while SPEX favors a more 
self-consistent approach. Most of the models that require atomic data in SPEX use 
the same atomic database and the same routines calculating the physical processes.

The intention is that the SPEX calculations are fast enough to run on an average
desktop computer, while retaining the physical information of the model. The 
aim of the code is to fit high-resolution X-ray spectra obtained by, for example, 
the Reflection Grating Spectrometer (RGS) aboard XMM-Newton \citep{denherder2001}
or the grating spectrometers aboard Chandra \citep{weisskopf2000}. With the
introduction of high-resolution microcalorimeter spectrometers on Hitomi 
\citep{takahashi2016}, XRISM\footnote{\url{http://xrism.isas.jaxa.jp/en/}} and 
ATHENA\footnote{\url{https://www.the-athena-x-ray-observatory.eu/}}, there will
be much higher requirements on the accuracy of the spectral models.

\section{Development Challenges}

The increasing spectral resolution of observed X-ray spectra poses a number of challenges.
Not only do we need to increase the size of the atomic database, also the accuracy
of the calculations needs to be improved. In addition, the new microcalorimeter 
instruments have much larger response matrices. These matrices are used to convolve 
the model spectrum with the instrument characteristics to make a direct comparison 
with observed spectra possible. Finally, there is an increasing demand to make the 
code better scriptable using languages like Python. At the same time, the code should 
remain light enough to run smoothly on a typical workstation.

\subsection{Dealing with More Atomic Data}

To illustrate the progress of the atomic database, we compare the number of lines 
used in the original SPEX version 2 models (before 2016) and the current version 
3 database. In version 2 models, the number of lines is of the order of 5000, 
while the current database contains about 2 million lines. Figure~\ref{fig:spex_compare}
shows the difference between these models around the silicon and sulfur lines 
around 2 keV. The bottleneck of calculating all these lines is to solve the line 
strengths (transition rates) per ion. Currently, this is done by solving a large 
set of rate equations using CPU accelerated LAPACK routines (Intel Math Kernel 
Library or OpenBLAS). An attempt to accelerate the calculation even more using 
graphical processors (GPU) did not result in faster execution times.   

\articlefigure{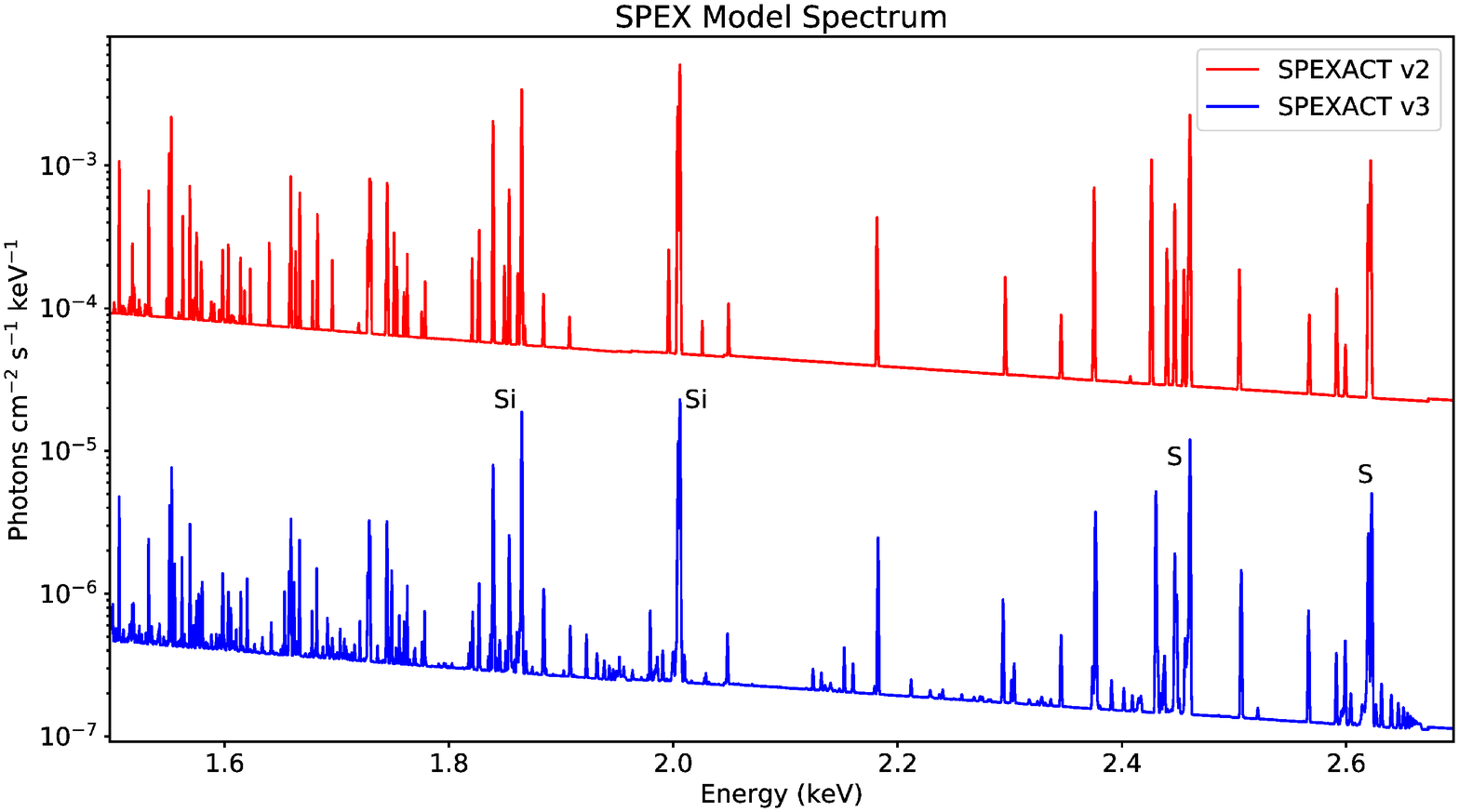}{fig:spex_compare}{Here we compare a thermal spectrum 
(with a temperature of kT = 1.5 keV) around the silicon and sulfur lines for SPEX 
versions 2 and 3.}

Another trick to speed up the calculations is to approximate computationally 
intensive physics relations with a simpler parametrization. It turns out that
some of these approximations in the SPEX version 2 models are not accurate 
enough for the high quality data we can expect from future X-ray missions.
Therefore, we are in the process of replacing the old functions with more 
accurate versions, if applicable also with more (recent) atomic data.

\subsection{Optimal Binning and Response Matrices}

Although model calculations are the most computationally intensive, we can also 
expect longer execution times when fitting high-resolution data. In X-ray 
astrophysics, it is customary to convolve the model spectrum with the instrumental
response (effective area and redistribution matrix) to calculate what the
model spectrum looks like when it is observed by the instrument. This way, 
the spectral model can be fitted to the observed spectrum.

With the increased spectral resolution of the new spectrometers, 
the redistribution matrices will become much larger. In the current standard 
OGIP format\footnote{\url{https://heasarc.gsfc.nasa.gov/docs/heasarc/ofwg/docs/spectra/ogip\_92\_007/ogip\_92\_007.html}},
these response matrices can grow to about 4 GB, while currently the largest 
matrices are of the order of 100 MB. To reduce this size, and
speed up the convolution, new response file structures and binning algorithms 
have been developed \citep{kaastra2016} to reduce the size of these matrices.
In addition, the matrices can be re-ordered to make it easier to do the convolution
in parallel using OpenMP. These features should help to keep the execution times for 
spectral fitting within reasonable limits.

\subsection{Creating Flexible Interfaces}

Currently, SPEX is an interactive program operated by giving commands on 
a command prompt in a terminal window. Although the interface can be quite 
efficient, once you learn the commands, there is an increasing demand from 
users to be able to script SPEX. Especially Python has grown very popular 
as a scripting language and together with Jupyter Notebook this provides very 
nice features to do a documented interactive analysis of spectra. We are 
currently developing a Python wrapper for SPEX that allows users to 
include SPEX into their Python program without using the command prompt.
If successful, this may appear as an experimental feature in the
release of SPEX 3.06.00.

In preparation for the Python interface, we released a package with 
Python versions of a few auxiliary tools\footnote{\url{https://github.com/spex-xray/pyspextools}}
\citep{deplaa2019}. This package is mainly focused on manipulating 
spectrum and response files in SPEX and OGIP format and contains a 
template to build user defined models for SPEX in Python.

\section{Code Distribution}

SPEX is written mostly in FORTRAN 90 and runs on UNIX systems like 
Linux and MacOS. Historically, SPEX has been distributed with statically compiled 
binaries included. Since SPEX version 3.05.00 the source code of SPEX is available
under the GPLv3 license following the new Open Science policy in the Netherlands.
This allows users to compile the code on their own machine, which solves issues 
with binary compatibility and library dependencies that sometimes occur at the cost
of a slightly more complicated install procedure. In addition to the binaries and 
source code, a Docker image\footnote{\url{https://hub.docker.com/r/spexxray/spex/}} is 
available to be able to run SPEX in a controlled environment. Since 2018, all 
SPEX versions above version 2.0 are available on Zenodo \citep{kaastra2018}.

\section{Discussion and Future Prospects}

Although a lot of work has been done to prepare SPEX for the Hitomi mission, there
are still challenges left on multiple aspects of the package. We intent to improve
the atomic data, expand and make the spectral models more accurate, improve the efficiency
of the spectral fitting, make the interfaces more flexible and user friendly, and
expand the available documentation with more examples and analysis threads.

The high-resolution spectra that will be obtained after the Japanese XRISM mission is 
launched early 2022 will provide thorough tests for the improvements made to the SPEX
package. This will help preparing SPEX and other spectral codes for 
the next generation of high-resolution spectrometers provided by ATHENA in the 2030's.

\bibliography{P11-15}

\begin{thebibliography}{}
\expandafter\ifx\csname natexlab\endcsname\relax\def\natexlab#1{#1}\fi
\expandafter\ifx\csname url\endcsname\relax
  \def\url#1{\texttt{#1}}\fi
\expandafter\ifx\csname urlprefix\endcsname\relax\def\urlprefix{URL }\fi
\providecommand{\eprint}[2][]{\url{#2}}

\bibitem[{{Arnaud}(1996)}]{arnaud1996}
{Arnaud}, K.~A. 1996, in Astronomical Data Analysis Software and Systems V,
  edited by G.~H. {Jacoby}, \& J.~{Barnes}, vol. 101 of Astronomical Society of
  the Pacific Conference Series, 17

\bibitem[{de~Plaa(2019)}]{deplaa2019}
de~Plaa, J. 2019, Pyspextools.
  \urlprefix\url{https://doi.org/10.5281/zenodo.3245804}

\bibitem[{{den Herder} et~al.(2001)}]{denherder2001}
{den Herder}, J.~W., et~al. 2001, \aap, 365, L7

\bibitem[{{Kaastra} \& {Bleeker}(2016)}]{kaastra2016}
{Kaastra}, J.~S., \& {Bleeker}, J.~A.~M. 2016, \aap, 587, A151.
  \eprint{1601.05309}

\bibitem[{{Kaastra} et~al.(1996){Kaastra}, {Mewe}, \&
  {Nieuwenhuijzen}}]{kaastra1996}
{Kaastra}, J.~S., {Mewe}, R., \& {Nieuwenhuijzen}, H. 1996, in UV and X-ray
  Spectroscopy of Astrophysical and Laboratory Plasmas, 411

\bibitem[{Kaastra et~al.(2018)Kaastra, Raassen, de~Plaa, \& Gu}]{kaastra2018}
Kaastra, J.~S., Raassen, A. J.~J., de~Plaa, J., \& Gu, L. 2018, Spex x-ray
  spectral fitting package.
  \urlprefix\url{https://doi.org/10.5281/zenodo.2419563}

\bibitem[{{Mewe}(1972)}]{mewe1972}
{Mewe}, R. 1972, \solphys, 22, 459

\bibitem[{{Takahashi} et~al.(2016)}]{takahashi2016}
{Takahashi}, T., et~al. 2016, in Space Telescopes and Instrumentation 2016:
  Ultraviolet to Gamma Ray, vol. 9905 of SPIE Proc., 99050U

\bibitem[{{Weisskopf} et~al.(2000){Weisskopf}, {Tananbaum}, {Van Speybroeck},
  \& {O'Dell}}]{weisskopf2000}
{Weisskopf}, M.~C., {Tananbaum}, H.~D., {Van Speybroeck}, L.~P., \& {O'Dell},
  S.~L. 2000, in SPIE Proc., edited by J.~E. {Truemper}, \& B.~{Aschenbach},
  vol. 4012 of Society of Photo-Optical Instrumentation Engineers (SPIE)
  Conference Series, 2. \eprint{astro-ph/0004127}

\end{thebibliography}


\end{document}